\documentclass[11pt, logo, nonumbering]{preprint}
\usepackage[utf8]{inputenc}
\usepackage{microtype}
\usepackage{graphicx}
\usepackage{amsmath}
\usepackage{booktabs}
\usepackage{xcolor}
\definecolor{mydarkblue}{rgb}{0,0.08,0.45}
\usepackage{url}
\usepackage{stfloats}
\usepackage{xspace}

\usepackage{xspace}
\usepackage{xcolor}

\usepackage{minted}
\definecolor{codebg}{rgb}{0.98,0.98,0.98}
\setminted{
  fontsize=\scriptsize,
  bgcolor=codebg,
  breaklines=true,
  tabsize=2,
  autogobble=true
}

\usepackage{booktabs}
\usepackage{multirow}
\usepackage{algorithm}
\usepackage{algpseudocode}
\usepackage{wrapfig}
\usepackage{subcaption}
\usepackage{placeins} 
\usepackage{float} 
\usepackage{fontawesome5}
\usepackage{tikz}
\usetikzlibrary{arrows.meta,positioning,shapes.geometric,fit,backgrounds}

\newcommand{\sysname}{\textsc{Ulysses}\xspace}
\newcommand{\benchname}{\textsc{WebExploitGym}\xspace}

\makeatletter
\@ifpackageloaded{tcolorbox}{}{\usepackage{tcolorbox}}
\makeatother
\tcbuselibrary{skins,breakable,listings,minted}

\newtcblisting{focuscode}[1]{
  listing only,
  listing engine=minted,
  minted language=#1,
  minted options={linenos=false,fontsize=\tiny,breaklines=true,autogobble=true},
  colback=codebg,
  colframe=red!60!black,
  boxrule=0.5pt,
  arc=1pt,
  left=3pt,right=3pt,top=2pt,bottom=2pt,
  enhanced,
}

\newtcolorbox{pilotbox}{
  colback=orange!8, colframe=orange!55!black, boxrule=0.4pt, arc=1pt,
  left=6pt, right=6pt, top=4pt, bottom=4pt,
  before skip=8pt, after skip=8pt,
  fontupper=\small,
}

\providecommand{\Description}[2][]{}

\usepackage[colorlinks=true,linkcolor=mydarkblue,citecolor=mydarkblue,filecolor=mydarkblue,urlcolor=mydarkblue]{hyperref}
\usepackage{cleveref}

\usepackage[style=numeric-comp,maxbibnames=3,minnames=1,backend=bibtex,natbib=true, doi=false,eprint=false,url=false]{biblatex}

\begin{document}

\title{\sysname{}: A Security Harness For Synthesizing Verifiable Web Exploits}

\author{
\begin{tabular}{cccc}
\textbf{Aditya Oke}$^{1,*}$ &
\textbf{André Vicente Duarte}$^{1,2}$ &
\textbf{Rui Melo}$^{1,3}$ &
\textbf{Shubham Gandhi}$^{1}$\\
\textbf{Nachiket Kotalwar}$^{1}$ &
\textbf{Charmi Khandor}$^{1}$ &
\textbf{Danqing Wang}$^{1}$ &
\textbf{Carolyn Rosé}$^{1}$\\
\multicolumn{4}{c}{\textbf{Lei Li}$^{1}$}\\
\end{tabular}\\[0.5em]
\textsuperscript{1}Carnegie Mellon University, Language Technologies Institute\\
\textsuperscript{2}INESC-ID, Instituto Superior Técnico\\
\textsuperscript{3}Faculdade de Engenharia do Porto\\
}


\correspondingauthor={Corresponding author: \href{mailto:aoke@cs.cmu.edu}{\nolinkurl{aoke@cs.cmu.edu}}} 

\maketitle

\begin{abstract}
How can an AI agent automatically discover security vulnerabilities in any software repository? Existing approaches use LLMs to generate software vulnerability exploits. However, it is challenging for an LLM to reliably verify its own generated exploits. In this paper, we propose \sysname{}, a multi-agent security harness to synthesize verifiable vulnerability exploits for web applications. \sysname{} converts candidate vulnerability findings into verified Proofs of Exploitability by iteratively exploring exploit hypotheses and establishing success in independently observed execution effects. We evaluated \sysname{} on \benchname{}, a curated benchmark of exploitable instances from 25 real-world web applications that span 5 languages and 7 frameworks. \sysname{} produces verified proofs for 10 of 25 real-world advisories. It also discovered 4 new proofs of exploitability that have never been previously reported.

\end{abstract}

\section{Introduction}

Security analysis often begins with finding vulnerabilities using static analysis tools, code reviews, or automated security review tools. However, a finding is ultimately a \emph{hypothesis} as to whether a code path is vulnerable. The strongest evidence for a vulnerability is a Proof of Exploitability (PoE) \cite{shakevsky2026antiproofsynthesizingvulnerabilitydetectors}. It demonstrates an attacker's capability against a deployed system. However, converting a finding into a verified exploit is difficult \cite{zhao2026anypocuniversalproofofconcepttest}. It requires understanding the complete attack path, application state, input transformations, authentication requirements, and the construction of a payload that successfully triggers the vulnerability.

Large language models (LLMs) are increasingly being used for vulnerability detection because they can reason about source code and application logic and identify vulnerability patterns that are difficult to capture with traditional rule-based analysis ~\cite{li2025iris, wang2026qlcoder}. The LLM-generated findings should still be treated as potential vulnerabilities and need actual confirmation before being triaged. The report does not establish that an attacker can successfully exploit the reported flaw. LLM Agents are automated and easier for developers to run, generating candidate findings at scale. This amplifies the existing triage burden by producing large volumes of potential but unverified findings.

\begin{figure}[!htbp]
    \centering
    \includegraphics[width=\linewidth, keepaspectratio]{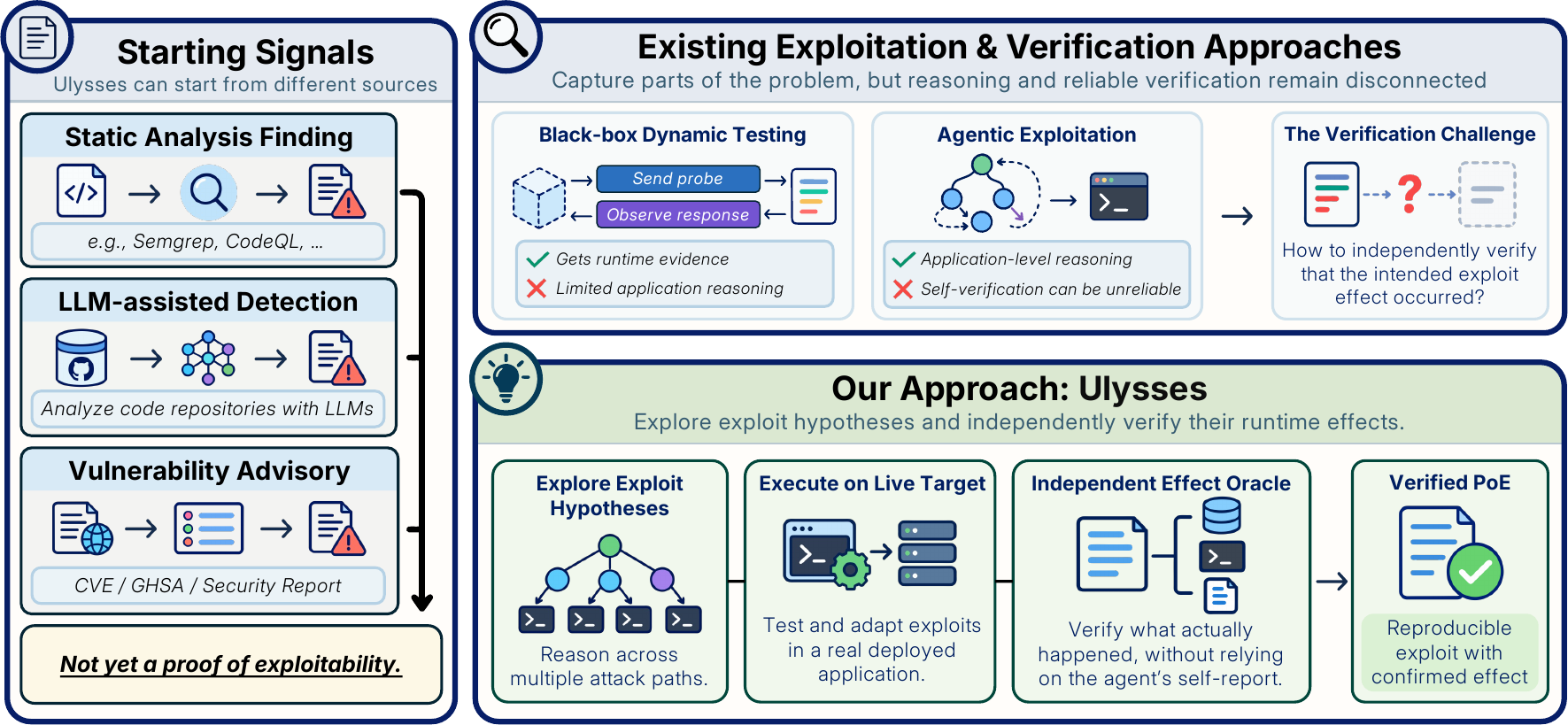}
    \caption{Black box approaches are limited by their methods, while agentic approaches rely on LLM self-judgment, making them unreliable. \sysname{} takes a different approach by exploring hypotheses and independently verifying runtime effects to create Proofs of Exploits (PoE).}
    \label{fig:crown}
\end{figure}

Recent experience with real-world security reporting illustrates this problem. Curl, one of the most widely used libraries for transferring files, reported that its public bug bounty program~\cite{curlBlogBounty} had about 20\% of vulnerabilities incorrectly generated by AI. These findings were overwhelmingly low quality, eventually leading to the shutdown of the program \cite{curlBountyClosed}. Although AI-generated reports were fluent and plausible, they lacked executable findings or proof that the flaw was real \cite{curlSlopVuln}. In May 2026, GitHub revised its bug bounty program to reduce the volume of low-effort and AI-generated reports \cite{ghrestructuring}. It suggests that developers focus on quality and provide proofs-of-concept that demonstrate real security or threat \cite{ghraisingbar}. Improving language models to generate vulnerability reports does not necessarily improve the ability to verify vulnerabilities. 

The problem is not limited to curl. Many real-world advisories on real-world repositories lack a complete proof of concept to trace, which makes a reported flaw hard to reproduce or debug. In May 2026 alone, the GitHub Advisory Database published 1,560 reviewed advisories, the highest in its history \cite{ghblog}. Sampling these advisories, we estimate that roughly half lacked Proof of Exploit. A developer reading such an advisory can see \emph{that} a flaw was reported but not \emph{whether} it can be exploited in their deployment.

There are two main challenges in producing the missing proof. First, it is difficult to distinguish an exploitable vulnerability from a candidate finding. An untrusted source is transformed several times before it reaches the vulnerable sink, and an exploit requires reasoning through the entire chain. The second is the verification challenge. Even after constructing a candidate exploit, it can be difficult to determine whether the vulnerability was actually triggered. In particular, relying on a model to judge its own exploit can result in incorrect claims of success without evidence from the application's execution.

A working PoE addresses both challenges by demonstrating the security impact of a reported finding. Rather than relying on the hypotheses from a vulnerability report or the model's own assessment, a PoE demonstrates that the vulnerable path can be reached and that the intended exploit succeeds under the application's execution conditions.
Based on this premise, we present \sysname{}, our automated PoE generation harness.

\paragraph{\textbf{Contributions:}}

\begin{itemize}

\item \textbf{\sysname{}: Open Harness for synthesizing verifiable proof of exploits (PoE)}

\sysname{} consumes the findings of any static analyzer, vulnerability finder, or advisory and produces the effect-verified proof of Exploitability (PoE). It also refines the findings into evidence-backed reports with higher precision.

\item \textbf{\benchname{}: Benchmark for Exploitable Web Applications}
We curate 25 instances that spans 5 languages and 7 frameworks of verifiable Web CVEs.

\item \textbf{Evaluation and Real World Impact}
We evaluated \sysname{} on 25 real-world applications and the OWASP Python benchmark in eight effect types. \sysname{} produces effect-verified PoEs for 10 of 25 advisories with advisory-informed localization, including working exploits for 4 advisories that previously had none.

\end{itemize}

\section{Background}
\label{sec:background}

\paragraph{\textbf{Web Applications and Vulnerabilities:}}
Billions of web applications are served online. Almost every internet user interacts with an application on the web. Web applications expose functionality over HTTP requests and accept input via request parameters, headers, and bodies. In source code, an externally controlled input enters the application at a \emph{source} and propagates through program logic before reaching a potentially security-sensitive operation, or \emph{sink}. A vulnerability arises when an untrusted input can reach a sensitive operation without adequate validation. This lets an attacker read or modify data, bypass authentication, or, at worst, execute arbitrary code. MITRE standardizes these flaws by describing the Common Weakness Enumeration (CWE) as a taxonomy of the types of underlying weaknesses. It publishes the Common Vulnerabilities and Exposures (CVE) list, which assigns an identifier to each vulnerability. A new vulnerability found is called a zero-day (Day 0), while once it is publicly disclosed, it is called an N-day (Day N) vulnerability.

Separately from this disclosure status, we describe three conditions based on how much an agent knows about a target vulnerability. In the \emph{advisory-informed}   condition, the agent receives the public facts an advisory carries, the weakness class, the affected component, and the intended effect, but no mechanism, payload, or proof. In the \emph{zero-knowledge} condition, the agent does not receive an advisory and must locate and exploit the flaw from the source code and the running endpoint. In the \emph{static-analysis-seeded} condition, the agent additionally receives the findings of a static analyzer over the source, which may or may not localize the vulnerability. All conditions target disclosed (N-day) vulnerabilities and differ only in what the agent is told. Zero-knowledge here refers only to the attacker's lack of prior information about the target.

\paragraph{\textbf{Application Exploits: }}

Many web vulnerabilities, when exploited, create an observable effect that can be validated. Injection flaws can run an attacker's command, read or modify database records, or send data to an external host. Broken access control can expose files or records of other users. Authentication and authorization failures can grant a session the attacker should not have, or even admin access. Path traversal can read protected files or write new ones. \sysname{} targets eight of such vulnerabilities: unauthorized file read, file creation, unauthorized db read, db modification, data exfiltration, remote code execution, privilege escalation, and authentication bypass. Each can be observed directly on the running system, such as a file being read, a database record returned, or a connection made back to the attacker.


\section{Motivation}

\begin{figure*}[t]
  \centering

  \begin{subfigure}[t]{0.49\textwidth}
    \caption{OWASP Benchmark Vulnerable Endpoint 164}
    \label{fig:intro-vuln}
    \input{fig/intro_code_snip}
  \end{subfigure}
  \hfill
  \begin{subfigure}[t]{0.49\textwidth}
    \caption{\sysname{} generated Proof of exploit (PoE)}
    \label{fig:intro-poc}
    \input{fig/intro_poc_patch_snip}
  \end{subfigure}

  \caption{\textbf{From vulnerability to verified exploit.} Left: Example from OWASP Python Benchmark shows an insecure deserialization endpoint that feeds sliced, attacker-controlled input to \texttt{pickle.loads}. Right: \sysname{}'s PoE, a pickle gadget that opens a reverse shell, verified by the callback to the oracle's listener.}
  \label{fig:intro-vuln-poc}
\end{figure*}

Figure~\ref {fig:intro-vuln-poc} (left) shows a vulnerable endpoint from the OWASP Python Benchmark~\cite{owasp-benchmark-python}. Static analysis tools~\cite{codeql,semgrep} mark the \texttt{pickle.loads} call as an insecure deserialization sink. Reaching this sink requires reasoning about previous steps and transformations. The request parameter is wrapped with \texttt{"help"+ param + "snapes on a plane"}, and the slice \texttt{[4:-17]} strips exactly the prefix and suffix. The attacker can control the bytes passed into \texttt{pickle.loads} and can supply a base64 URL-safe object to reach the sink. The attacker must recognize this transformation and construct a serialized object that triggers the desired behavior.

Figure ~\ref{fig:intro-vuln-poc} (right) shows the PoE produced by \sysname{}. \sysname{} generates an exploit  whose \texttt{\_\_reduce\_\_} method invokes \texttt{os.system} with a reverse-shell command, encodes it with base64 URL-safe encoding and submits it as the parameter.  When the endpoint deserializes the payload, it executes the command and connects a shell back to the oracle's listener. The incoming connection allows the oracle to independently confirm the exploit, helping \sysname{} generate accurate, verifiable Proof of Exploit (PoE).

Real-world application vulnerabilities are more complex. Applications often span multiple files, making it harder to locate a vulnerable data flow. Identifying a possible source and sink is only a starting point. An exploit must be constructed by reasoning through the intermediate transformations needed to carry attacker-controlled input to the sink and trigger the intended behavior.

Naturally, when exploiting, a security expert would like to determine whether the exploit achieved its intended effect. However, the appropriate way to verify success depends on the type of exploit. For example, after gaining Remote Code Execution (RCE) access, an expert can verify the result by opening a shell or forwarding a command to the target. Similarly, for a file read vulnerability, successful exploitation can be verified by reading secret data from the target. This notion of using observable evidence to verify task completion is closely related to Capture-The-Flag (CTF) style tasks~\cite{shao2024nyuctfbench}, where LLM agents operate in black-box environments and use hidden flags as evidence to show task completion. \sysname{} introduces planted, observable effects that serve as evidence of successful exploitation to produce the proof of exploits.
\section{Design}

The design of \sysname{} is guided by four challenges in PoE generation. First, vulnerability discovery and exploit verification are different tasks, since finding potentially vulnerable code does not establish that the finding is exploitable. Secondly, a finding may also have multiple possible exploit paths, making it necessary to explore different hypotheses rather than following a single path. Third, complex vulnerabilities require chaining multiple exploit steps. This requires the system to plan and adapt from previous attempts. Finally, an exploiting agent cannot reliably verify its own success. In the following sections, we describe these challenges and the corresponding mechanisms used by \sysname{}.

\begin{figure}[!htbp]
    \centering
    \includegraphics[width=1\linewidth]{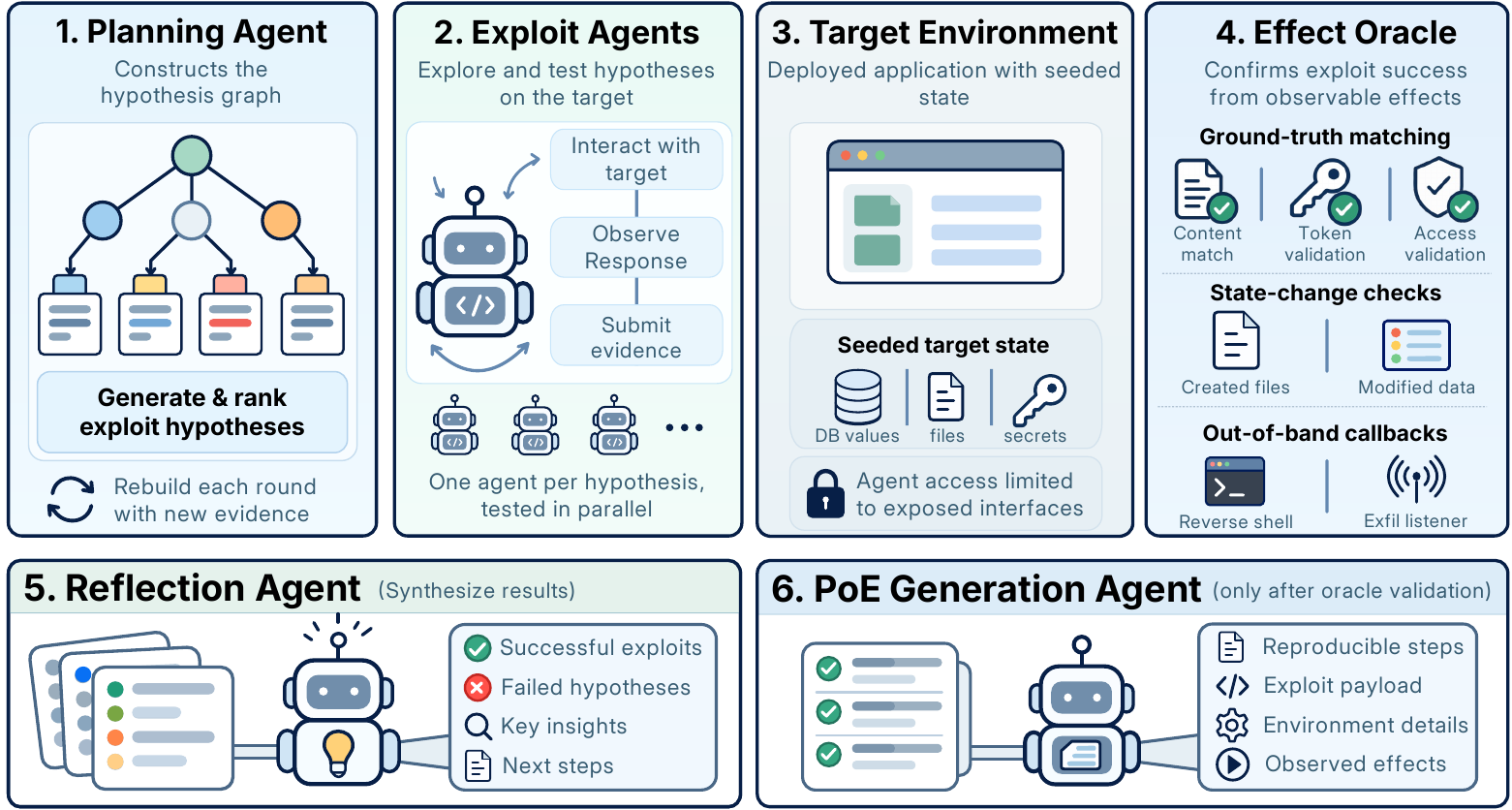}
    \caption{Exploit Agent Architecture}
    \label{fig:arch}
\end{figure}

\subsection{Vulnerability Scanning}
\label{sec:scanning}

Vulnerability detection involves the use of tools for initial surfacing and multi-step analysis. These could be rule-based static analyzers such as Semgrep, CodeQL, or Bandit ~\cite{semgrep, codeql, bandit}, an LLM-assisted scanner such as Absentia~\cite{duarte2026absentiadetectingbrokenaccess} or IRIS ~\cite{li2025iris}, or even a bare CVE description with no candidate location at all. These tools specialize in the detection and marking of sources and sinks, and providing attack plans or information about possible vulnerable locations. \sysname{} relies on the use of the above information as a starting point for exploits. Separation of detection from exploitation allows \sysname{} to focus on verifying the findings to report real exploitable vulnerabilities.

\subsection{Hypothesis Graphs}
\label{sec: hypothesis graph}

The Planning Agent proposes candidate exploits as a hypothesis graph, where each node is a candidate exploit, and each edge carries the context and reasoning needed to reach the sink. \sysname{} chose graph-based execution over a single linear agent loop for three reasons. The first is that an application's attack surface rarely reduces to one path. Several candidate exploits are usually worth pursuing at once, and a single agent reasoning about all of them serially would need to hold their combined context. The second is that exploitation is not one-shot. A hypothesis that fails in one round may become viable once a later round discovers a different entry point, so the graph is rebuilt each round from the accumulated findings and the previous round's reflection. This lets later rounds pursue paths that earlier rounds missed. Finally, graph-based execution keeps context passing efficient. Each node receives only the context carried by the edges leading into it, rather than the full history of everything attempted so far. Finally, using hypothesis graphs \sysname{} avoids the periodic compaction that a single long-running agent loop would otherwise need to remain within its context window. ~\cite{scalingagents}.

\subsection{Effect Oracle}
\label{sec:effect oracle}

The effect oracle acts as an independent proxy for the real application, checking whether an attempted exploit actually succeeded. The agent has no direct access to the target's database, file system, or operating system, so it cannot confirm success the way an attacker with real access could. The oracle can, because it alone holds the ground truth it planted, such as a seeded file, an in-scope database value, or a signing key, and reports back only whether the submitted evidence matches that ground truth, without revealing the value itself.

\begin{table*}[h]
    \captionsetup{justification=raggedright,singlelinecheck=false}
    \raggedright
    \caption{The eight effect types: their representative weakness class, the evidence the agent produces, and how the oracle validates each.}
    \label{tab:effects}
    \begin{tabular}{llll}
      \toprule
      \textbf{Effect Type} & \textbf{Agent Evidence} & \textbf{Oracle Check} \\
      \midrule
      Unauthorized File Read & File contents & Content matches planted secret \\
      Unauthorized File Creation & Created file path & Seeded file present in target \\
      Database Read & Extracted values & Value matches an in-scope DB row \\
      Database Modification & Modified value & Scoped row changed to target value \\
      Authentication Bypass & Session token & Token verifies against runtime secret \\
      Privilege Escalation & Admin session token & Admin-only endpoint reached with token \\
      Data Exfiltration & Beacon with payload & Seeded payload reaches exfil listener \\
      Remote Code Execution & Reverse shell & Reverse shell reaches RCE listener \\
      \bottomrule
    \end{tabular}
\end{table*}

A submission might fail because a token's subject claim is wrong for an authentication bypass or because the submitted content does not match the seeded file for a file read. The oracle responses help the agent refine its next attempt rather than repeat the same mistake. Table~\ref{tab:effects} lists the eight types of effects that \sysname{}  verifies, the evidence that a candidate must exploit, and what the oracle checks against. The oracle is hardened and cannot be satisfied by guessing values or brute-force attempts. We harden it by placing randomly generated flags in the oracle check mechanism described in Table ~\ref{tab:effects}. The seeded flags need to be discovered by exploiting the application. The effect oracle verifies whether the flags match. The proof of exploit is generated by \sysname{} only when the effect oracle confirms the correct flag submission.


\sysname{} runs on two separate hosts interconnected by a virtual network. The agent runs on Host A inside a Docker sandbox. The target application, the effect oracle, and the effect listeners run on Host B. The target is served as an HTTP Web server running as a Docker container. The Effect oracle has two listeners, one for RCE and the other for data exfiltration. The security group controls what ports can be reached on Host A and Host B. We allow the agent to reach the target and the effect oracle. To prevent false submissions, the agent is denied access to listener ports. Only Host B can reach its own listeners. This ensures that a reverse shell or a beacon on a listener can come only from the target.

\section{Approach}
\label{sec:method}

\sysname{} runs a loop of planning, exploitation, and reflection on the hypothesis graph. The loop runs for several rounds until a time budget or a round limit is reached. The figure \ref{fig:arch} describes our overall flow.

\subsection{Threat Model}

The agent runs as an attacker. It has white-box access to the source code and reaches the hosted target over the network. It has no direct access to the target's database, file system, or operating system. The oracle is an independent arbitrator. It knows only the ground truth it planted, such as the names of seeded files, in-scope database values, or the runtime signing key, and it does not know which findings are exploitable. The agent and the target run on separate hosts, and the agent cannot reach the oracle's effect listeners, so it cannot fabricate a callback when submitting a proof. Since the agent never sees the planted seeds, it cannot forge evidence of exploits.

\subsection{Agent Loop}
Algorithm ~\ref{alg:loop} describes the loop. The Planning Agent proposes hypotheses. \sysname{} tests every hypothesis in parallel and marks one as exploitable when the oracle validates it. The Reflection Agent then summarizes the round and seeds the next planner. Every agent has a turn limit to avoid stalling.  The loop returns the refined report and the effect-verified PoEs.

\begin{algorithm}[h]
  \caption{\sysname{} Agent loop}
  \label{alg:loop}
  \begin{algorithmic}[1]
  \Require findings $F$, target $T$, max rounds $R_{\max}$
  \Ensure refined report $R$, oracle-verified PoEs $P$
  \State $\textit{reflection} \gets \bot$;\quad $A \gets \emptyset$;\quad $r \gets 0$
     \Comment{$A$: cumulative results}
  \While{$r < R_{\max}$}
    \State $r \gets r + 1$
    \State $H \gets \textsc{PlannerAgent}(F, T, \textit{reflection})$
       \Comment{profile + source + scope + last reflection}
    \For{\textbf{each} $h \in H$ \textbf{in parallel}}
       \State $x_h \gets \textsc{ExploitAgent}(h, T)$
          \Comment{attempts exploit; may submit proof to the oracle}
       \If{$x_h.\textit{status} \neq \texttt{exploitable}$ \textbf{and}
           $\textsc{OracleValidated}(x_h)$}
          \State $x_h.\textit{status} \gets \texttt{exploitable}$
          \Comment{oracle override}
       \EndIf
       \State $p_h \gets \textsc{PoEAgent}(h, x_h)$
              \textbf{ if } $\textsc{OracleValidated}(x_h)$ \textbf{ else } $\bot$
    \EndFor
    \State $A \gets A \cup \{(h, x_h, p_h) : h \in H\}$
    \State $\textit{reflection} \gets \textsc{ReflectionAgent}(\{x_h : h \in H\})$
       \Comment{round results $\rightarrow$ next planner}
    \State $R \gets \{\, x.\textsc{toFinding}();  \forall x \in A,\
                       x.\textit{status}{=}\texttt{exploitable} \,\}$
    \State $\textsc{Persist}(R)$ \Comment{refined report after each round}
  \EndWhile
  \State $P \gets \{\, p_h : \forall p_h \in A,\ p_h \neq \bot \,\}$
  \State \Return $R, P$
  \end{algorithmic}
\end{algorithm}

\subsection{Planning}
The Planning Agent reads the vulnerability scanner's attack plan and chains, the source, and the previous round's reflection. The planner is a broader analyzer and is more concerned with finding all possible scenarios that can lead to hypotheses. These include chaining attack chains or exploiting even low-probability ones. We test hypotheses in batches in order of decreasing confidence, as assigned by the planner. Each round, the planner tries to generate unique hypotheses. The reflection of the previous round guides it to generate plans that are more likely to succeed than the previous ones.



\subsection{Exploitation}
The Exploit Node creates a subagent for every hypothesis node. The exploiter node uses Python scripts or bash commands to test every hypothesis and trigger HTTP Requests against the target app. It also calls for the proof submitter to verify and submit proof of exploit to the effect oracle. The agent never judges the exploit by itself; instead, it triggers the payload and lets the oracle decide.  Once an exploit is validated by the effect oracle, the outputs are passed to the Proof of Exploit Generation Node. At the end of exploitation, a hypothesis is either exploitable, vulnerable but non-exploitable, or benign. We filter out vulnerable and benign reports and keep only confirmed exploits.

\subsection{Proof of Exploit Generation}

For every exploit the oracle verifies, a PoE node generates a Proof of Exploit. The PoE node is a separate agent, attached to the exploit node by an edge. It is a conditional node that fires only when the effect oracle signals exploit success. The PoE is a standalone Python script that reproduces the exploit on the deployed target. To evaluate PoE quality, we re-run them against the deployed targets and verify whether the expected effect is achieved.

\subsection{Reflection}

At the end of the round, the Reflection Agent analyzes the trajectories. It records the types validated so far and the types still uncovered. Reflection helps in prioritizing and creating feedback on what worked and what did not work.
This analysis becomes the planner's input for the next round. This helps \sysname{} improve the attempts in the next round and build multi-step chains across rounds.

\section{Benchmark Curation \benchname{}}
\label{sec:dataset}


\sysname{} verifies an exploit by observing its effect on a live target. Evaluating it requires a deployed and running web application and the source code at the vulnerable commit. Existing benchmarks target memory-safety bugs, build-only libraries, or synthetic code, and validate success by crash or by label rather than by executing a real exploit against a running web application. We curate \benchname{}, a benchmark of real web application CVEs, each packaged as a deployable instance with the seeded state needed to observe a successful exploit.

\begin{table*}[h]
\caption{\benchname{} compared to existing exploitation benchmarks.}
\label{tab:benchmarks}
\begin{tabular}{lllll}
  \toprule
  Benchmark & Domain & Target & Validation & Size \\
  \midrule
  \textbf{\benchname{} (ours)} & Web apps & Deployed & PoE execution & 25 \\
  OWASP Benchmark & Web apps & Deployed & Ground-truth labels & 1230 \\
  CWE-Bench-Java & Java libraries & Build-only & Patch Differential & 120 \\
  BountyBench & Web + libraries & Deployed & Differential checker & 40 \\
  CyberGym & C/C++ systems & Build + fuzz & Sanitizer Crash & 1507 \\
  CyberGym-E2E & C/C++ systems & Build + fuzz & Crash + patch & 920 \\
  ExploitBench & JS engine (v8) & Build + run & Crash & 41 \\
  \bottomrule
\end{tabular}
\end{table*}

\paragraph{\textbf{Data Collection}}

\benchname{} is built from vulnerabilities disclosed through the National Vulnerability Database (NVD) \cite{nist_nvd} and GitHub Security Advisories (GHSA) \cite{ghsa}. We keep web application vulnerabilities whose exploitation produces one of the eight observable effects \ref{sec:effect oracle}. Each data point ships with a deployable version of the affected application at its vulnerable release, together with the seeded values intended to observe the effect. It receives only the CVE identifier, a summary, and any credentials that the attacker needs. Submissions are scored by executing the agent's PoE against the deployed instance and confirming the intended effect. All CVEs are disclosed after the training cutoff of the models we evaluate. \benchname{} comprises 25 vulnerability instances across five programming languages and a diverse set of CWE weakness classes.

The full list of vulnerability instances included in our benchmark is provided in Appendix~\ref{sec:appendix} (Table~\ref{tab:instances}).

\paragraph{\textbf{Statistics}}

\benchname{} contains $25$ instances in $5$ languages (Go, Java, JavaScript, PHP, and Python) and $7$ web frameworks (including Laravel, Spring~Boot, Flask, FastAPI, Express, CakePHP, and PocketBase), covering the eight effect types, with CVEs disclosed between $2025$ and $2026$. $14$ instances have a public Proof of Exploit, while the remaining $11$ do not.  We choose models with training cutoff dates after the CVE was disclosed. The availability of a public exploit does not give the agent an informational advantage. All instances have a public NVD entry or a GitHub Security Advisory.

\section{Experiments}

\paragraph{\textbf{Implementation:}} 

\begin{figure}[htbp]
\centering
    
    \begin{subfigure}[t]{0.49\linewidth}
        \centering
        \includegraphics[width=\linewidth]{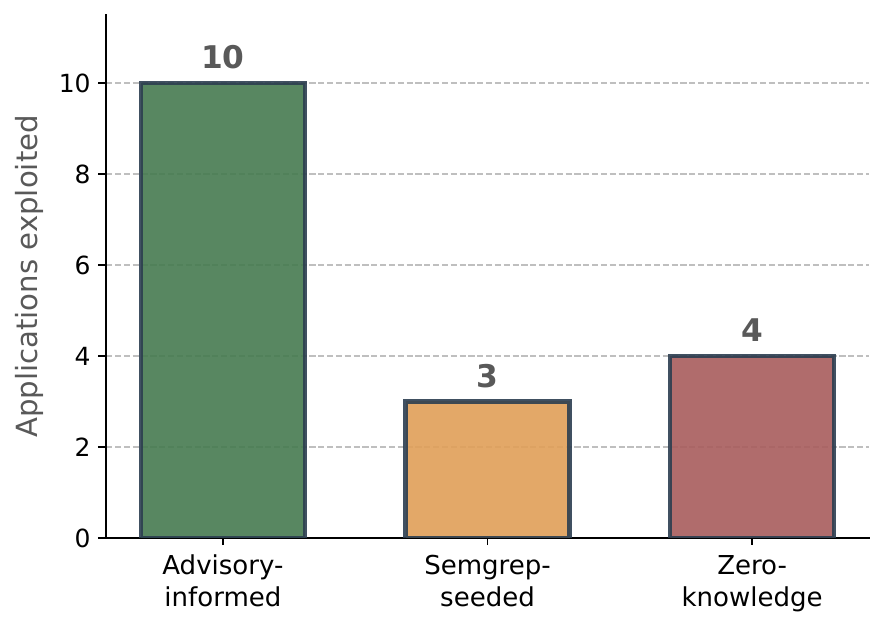}
        \Description{Vulnerability detection affects \sysname{} performance}
        \caption{Conversion of detected candidates into PoEs}
        \label{fig:baseline-breakdown}
    \end{subfigure}
    \hfill
    \begin{subfigure}[t]{0.49\linewidth}
        \centering
        \includegraphics[width=\linewidth]{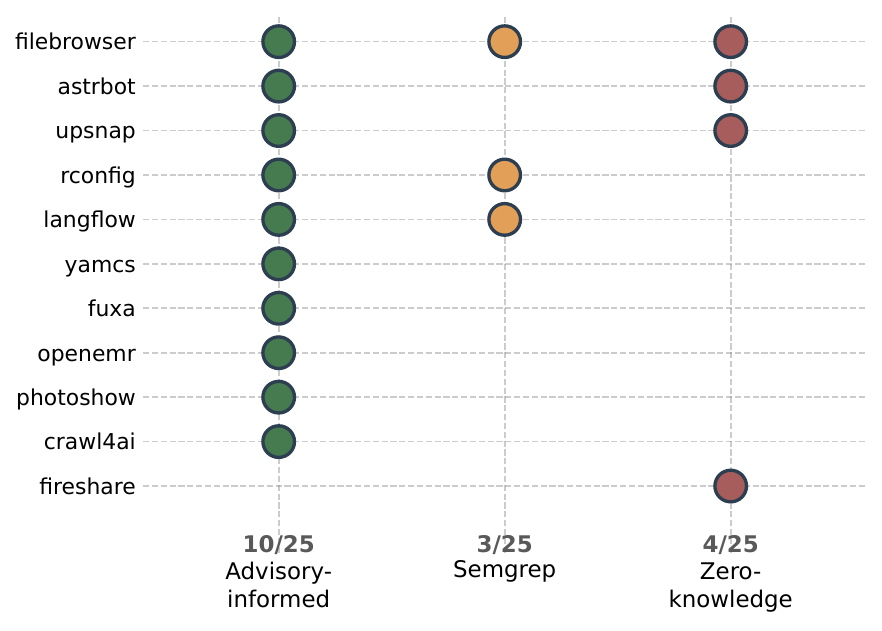}
        \caption{\benchname{} Exploited web applications}
        \label{fig:baseline-recall-fpr}
    \end{subfigure}

  \caption{\textbf{\sysname{} is effective at crafting PoEs when vulnerabilities are detected}}
  \label{fig:rq3}
\end{figure}

The planner is preferably a stronger model, since this is the stage where most cyber capabilities are needed. This stage also has a high probability of rejection from cyber-safeguarded models such as Claude or GPT \cite{cybersafeguards}. We use GLM-5 \cite{glm5team2026glm5vibecodingagentic}, an open-weight model available on Amazon Bedrock.
Additionally, this step is highly agentic and requires multiple tool calls in a limited time. We ablate the choice of exploit node models in  the architecture ablation Figure \ref{fig:rq3}. The PoE Generating model is kept the same as the exploit node. The reflection model is the same as the planner.

\paragraph{\textbf{Benchmarks:}} We evaluate the OWASP Benchmark for Python ~\cite{owasp-benchmark-python}, which are synthetic and carry ground-truth labels, and on \benchname{}, our dataset of real web CVEs. OWASP BenchmarkPython contains 1,230 test cases in total. We score against the 339 test cases across the six effect-verifiable categories (path traversal, command injection, SQL injection, code injection, insecure deserialization, XXE), 129 of which are ground-truth positive. \benchname{} is scored by executing the PoE. Findings only count if the generated PoE produces the desired effect.

\paragraph{\textbf{Baselines:}} For Vulnerability detection, we run five static analyzers, CodeQL \cite{codeql}, Semgrep \cite{semgrep}, Bandit \cite{bandit}, Bearer \cite{bearer}, and Joern \cite{yamaguchi2014joern}. In addition, we also tested the black box with the OWASP ZAP \cite{owaspzap} dynamic analyzer. Additionally, we also compared the use of the LLM-based Absentia method \cite{duarte2026absentiadetectingbrokenaccess} as a vulnerability detector.

\paragraph{\textbf{RQ1}}: \textbf{How effective is \sysname{} in converting detected vulnerabilities into PoEs ?}

\begin{figure}[htbp]
\centering
    
    \begin{subfigure}[t]{0.49\linewidth}
        \centering
        \includegraphics[width=\linewidth]{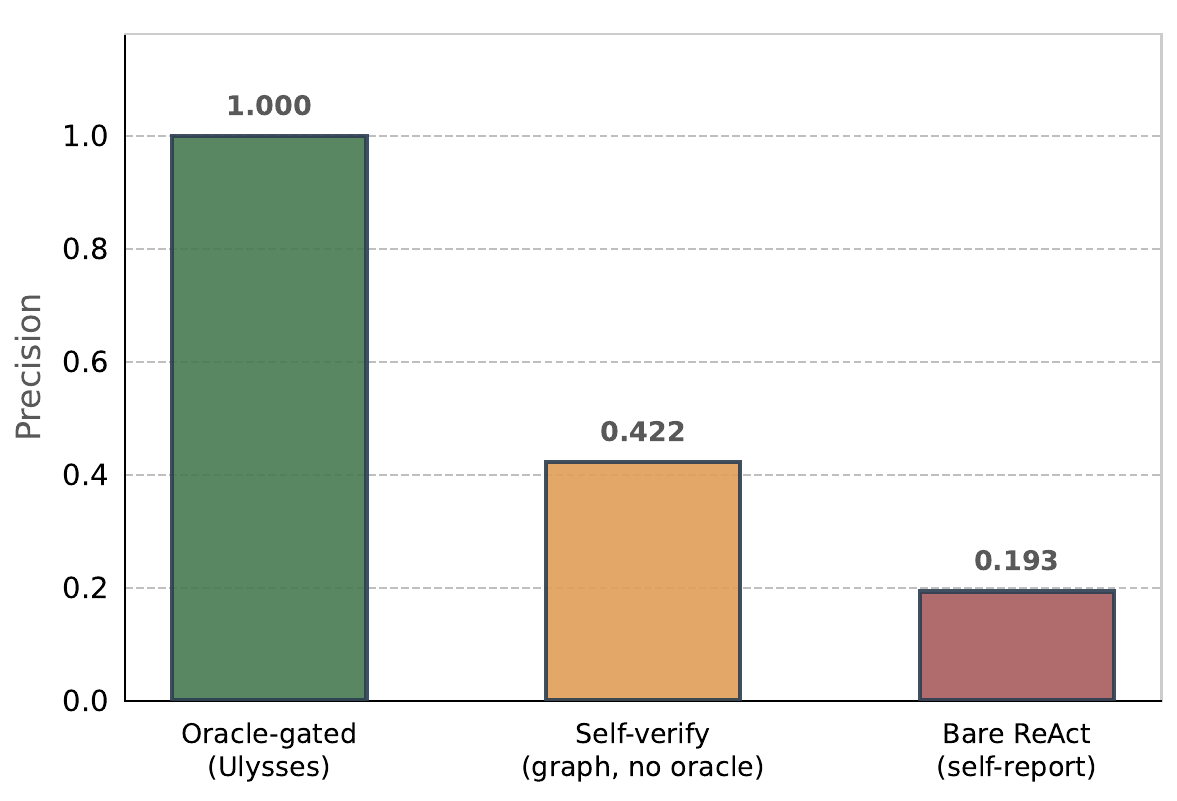}
        \Description{A plot showing the precision of self-reporting agents, demonstrating unreliability.}
        \caption{Self-report is not reliable. Precision falls for agents to confirm it's exploits}
        \label{fig:baseline-breakdown}
    \end{subfigure}
    \hfill
    \begin{subfigure}[t]{0.49\linewidth}
        \centering
        \includegraphics[width=\linewidth]{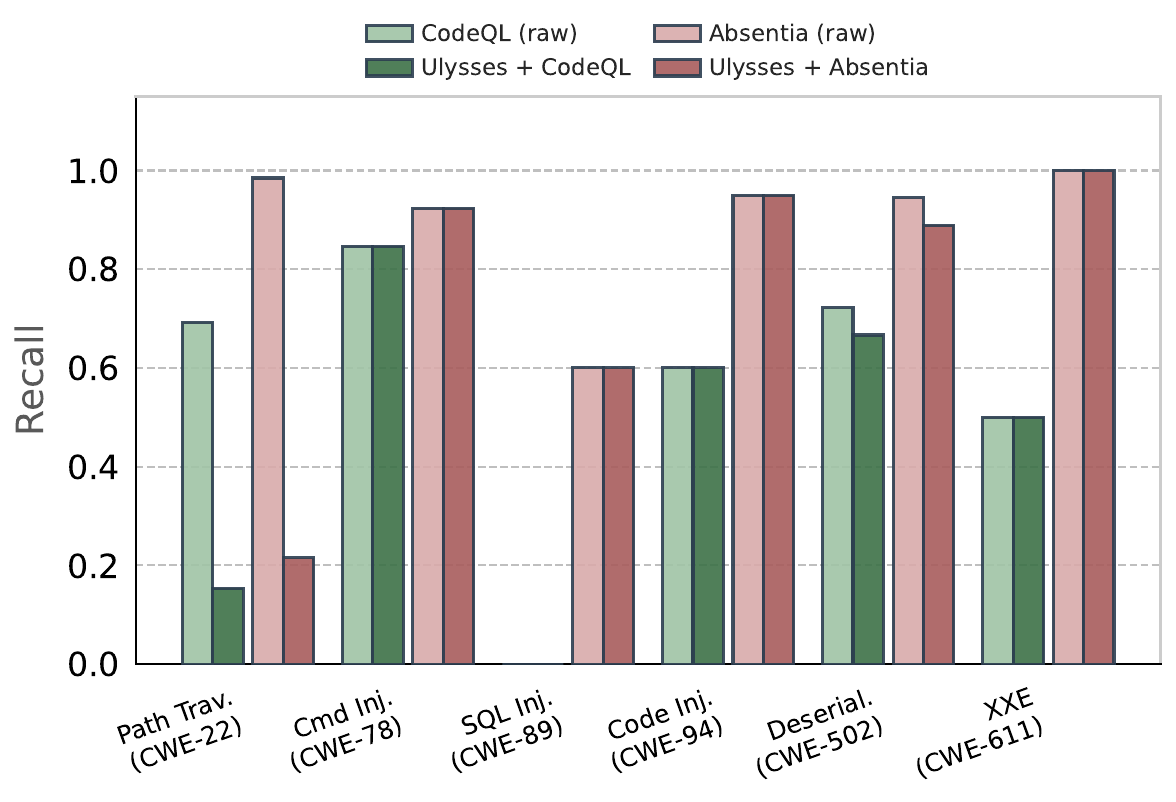}
        \caption{\sysname{} retains recall for 5 out of 6 CVE classes using effect oracle}
        \label{fig:baseline-recall-fpr}
    \end{subfigure}

  \caption{\textbf{\sysname{} converts vulnerabilities with high precision at low recall drop}}
  \label{fig:rq1}
\end{figure}

Converting candidate findings into reliable Proof of Exploitation (PoE) has two challenges. First, static analyzers are noisy. In Figure \ref{fig:rq1}, we analyze the injection and exploitable categories of OWASP BenchmarkPython (CWE 22/78/89/94/502/611); CodeQL flags candidates with 48\% precision and Absentia flags with 54\% precision. Roughly half of every flagged candidate is a false positive. Second, the agent's ability to self-judge exploits is weak. Pooled across both seed detectors and all three backbone models, precision falls to 0.422 for the full multi-hypothesis graph and to 0.193 for a simple single-loop ReAct agent using identical tools, against 1.000 by construction for the effect oracle pipeline.

Sophisticated agents fail less because they explore more hypotheses to test before deciding to report. Agents that reason about the exploit are the least reliable judges of whether the exploit actually worked. This is because of high faith in their evidence and reasoning process, while the effect is unobservable. \sysname{} fixes self-reporting issues by using the effect oracle to observe the exploit. Therefore, it never emits Proof of Exploit from an agent's self-report. It synthesizes only when the effect oracle has confirmed. 

\begin{pilotbox}
    \textbf{Takeaway}: \sysname{} converts noisy static-analysis output into verified exploits. An independent effect oracle, not a stronger detector or a more confident model, improves conversion.
\end{pilotbox}

\paragraph{\textbf{RQ2}}: \textbf{What drives \sysname{}'s performance?}

We ablate three independent factors while keeping the others fixed on the OWASP Python Benchmark dataset. 

\paragraph{\textbf{Effect Oracle:}} For the following experiments, we keep the effect oracle on and vary only the architecture, model, and vulnerability scanner.

\begin{figure}[htbp]
\centering
  \begin{subfigure}[t]{0.49\linewidth}
      \centering
      \includegraphics[width=\linewidth]{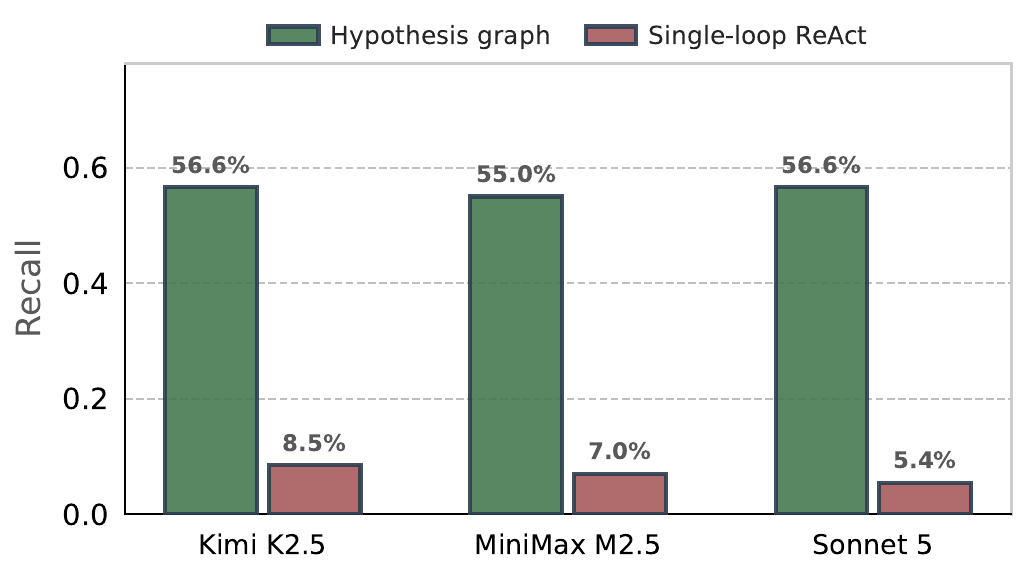}
      \Description{Grouped bar chart showing recall for the hypothesis graph versus a single-loop ReAct agent, at each of the three backbone models.}
      \caption{Architecture drives recall at every model.}
      \label{fig:rq2-a}
  \end{subfigure}
  \hfill
  \begin{subfigure}[t]{0.49\linewidth}
      \centering
      \includegraphics[width=\linewidth]{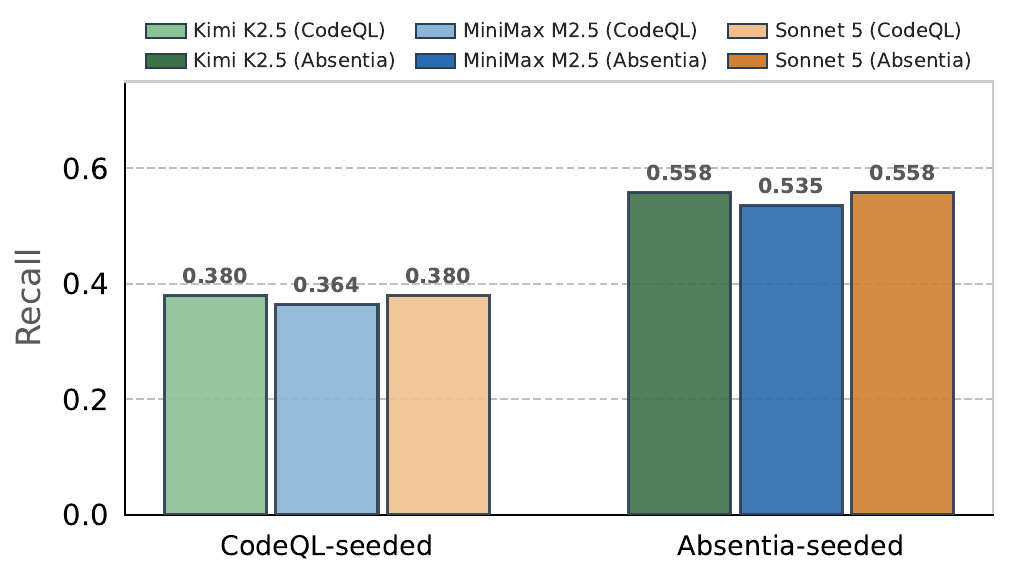}
      \Description{Grouped bar chart showing recall by seed detector, CodeQL versus Absentia, with one bar per model.}
      \caption{Recall is bounded by the vulnerability scanner.}
      \label{fig:rq2-b}
  \end{subfigure}
\caption{\textbf{Graph architecture and vulnerability scanner improve \sysname{} recall}}
\label{fig:rq2}
\end{figure}

\paragraph{\textbf{Architecture:}}
Figure \ref{fig:rq2} shows runs with hypothesis graphs and a ReAct-based agent with 3 different families of models. The Hypotheses graph and the ReAct Agent have the same 300-second budget per route. Sonnet~5, Kimi-K2.5, and Minimax-M2.5 land within 1.6 points of each other. The architecture choice matters more than the intrinsic capabilities of the model. Hypothesis graphs have much higher recall.

\paragraph{\textbf{Vulnerability Scanner:}} \sysname{} relies on the initial vulnerability scanner for the source attack surfaces. The scanner quality sets an upper bound on the achievable recall, regardless of the model or the verification loop. Absentia provides stronger potential candidates than CodeQL.

\begin{pilotbox}
    \textbf{Takeaway:} Three Factors account for \sysname{} improvements. The hypothesis-graph architecture improves recall. The seed bounds the recall improvements.
\end{pilotbox}

\paragraph{\textbf{RQ3}}: \textbf{Does \sysname{} exploit complex vulnerabilities in real-world web applications?}

We evaluated \sysname{} on \benchname{}, a corpus of real CVEs from web applications disclosed after the model training cutoff. Each target is deployed from its original source, with its public proof-of-concept withheld from the agent. The effect oracle confirms the exploit during the run, and we re-execute the generated PoE against the deployed target to confirm it reproduces. We report the conversion rate of the scaffolded targets on which \sysname{} produces an effect-verified exploit. We run the three information conditions defined in Section~\ref{sec:background}, advisory-informed, Semgrep seeded, and zero-knowledge, holding the model and the target fixed.

\sysname{} produces effect-verified PoEs for 10 of 25 targets under the informed advisory condition, compared to 4 of 25 under zero knowledge. Semgrep-seeded analysis identifies 3 targets, but the agent must still triage noisy findings to determine reachability, resulting in fewer successful conversions than when the advisory directly identifies the affected component.

\begin{pilotbox}
    \textbf{Takeaway:} The gap between informed and zero-knowledge conditions shows that localization, rather than verification, is the primary bottleneck. Given a location, \sysname{} reliably converts it into a verified exploit, while without one, it spends much of its budget on discovery.
\end{pilotbox}

\section{Case Study}

We use an instance from \benchname{} to illustrate a Proof of Exploit (PoE) using \sysname{}. We consider CVE-2026-77914 in rConfig~\cite{rconfig}, an 
application built on Laravel. 
The vulnerability is an authenticated path traversal in the export-download endpoint. 
The advisory (GHSA-m5rw-jcrm-mmwc) describes the vulnerability and affected component, but does not provide a PoE, reproduction steps, or a payload. 

\begin{figure*}[t]
    \begin{subfigure}[t]{0.49\textwidth}
      \begin{minted}[fontsize=\small,startinline=true]{php}
    // FileDownloadController.php rConfig
    // Route: GET /download-export
    public function download_export() 
    {
      // VULN: no basename()
      $path = export_path() . $_GET['filename'];
      if (file_exists($path)) {
          activityLogIt(__CLASS__, __FUNCTION__,
              'info',
              'File download: ' . basename($path),
              'downloader');
          return response()->download($path);
      }
      return \Response::json([
          'error'   => 404,
          'message' => basename($path)
                       . ' could not be downloaded'
      ]);
    }
      \end{minted}
      \caption{\textbf{Vulnerable \texttt{download\_export} endpoint in rConfig~8.2.12.} 
      The endpoint constructs the requested file path by concatenating the application-controlled export directory with the user-supplied \texttt{filename} parameter. 
      The input is not sanitized with \texttt{basename()} nor otherwise constrained to the export directory.
      An attacker can supply path-traversal sequences to access files outside the intended directory.}
      \label{fig:vuln-code}
    \end{subfigure}
    \hfill
    \begin{subfigure}[t]{0.49\textwidth}
      \begin{minted}[fontsize=\small]{python}
    import re, requests
    target = "http://172.31.37.38:8352"
    session = requests.Session()
    
    # Step 1: Authenticate (Laravel CSRF)
    resp = session.get(f"{target}/login")
    csrf = re.search(
      r'name="csrf-token" content="([^"]+)"',
      resp.text, re.IGNORECASE).group(1)
      
    session.post(f"{target}/login", data={
      "username": "admin@domain.com",
      "password": "admin", "_token": csrf},
      allow_redirects=False)
    
    # Step 2: Path traversal on download-export
    tr = "/../../../../../../../tmp/flag.txt"
    resp = session.get(
      f"{target}/download-export"
      f"?filename={tr}")
    # Step 3: Confirm file read
    assert "FLAG_" in resp.text
    print(f"[+] File content: {resp.text}")
      \end{minted}
      \caption{\textbf{\sysname{}-generated proof of exploit (PoE) for the path-traversal vulnerability.} 
      The generated script first authenticates to the rConfig instance using the application's login and CSRF mechanism, then supplies a traversal payload to the \texttt{/download-export} endpoint that escapes the export directory and targets \texttt{/tmp/flag.txt}.}
      \label{fig:poe-code}
    \end{subfigure}
    \caption{Case study of vulnerability exploitation in rConfig~8.2.12. The left panel shows the vulnerable download endpoint. The right panel shows the proof of exploit generated by \sysname{}.}
    \Description{Case study of a path-traversal vulnerability in rConfig~8.2.12. The left panel shows a download endpoint that concatenates user-controlled input with an export directory without sanitization. The right panel shows an agent-generated Python proof of exploit that authenticates to the application, submits a path-traversal payload targeting a file outside the export directory, and verifies that the file contents are returned.}
    \label{fig:case-study-vuln-poe}
\end{figure*}

\paragraph{\textbf{What makes this target difficult}}

The exploit requires multiple steps to work together. The agent must first authenticate through a CSRF-protected login flow and then identify the correct download endpoint and parameter, \texttt{filename}. The surrounding code also contains a \texttt{PathContainmentService} that protects other controllers. This makes it easy to incorrectly assume that the vulnerable endpoint is similarly protected.

\paragraph{\textbf{Zero Knowledge Scenario}}

In the zero-knowledge condition, \sysname{} does not receive any vulnerability-specific information. It must first identify that a vulnerability exists, localize it to the relevant code and endpoint, and then construct and verify the PoE. In this case, the planner eventually identifies the vulnerable source code, but is misled by the \texttt{PathContainmentService} and fails to establish that the vulnerable endpoint is actually reachable. Without a localization signal, the exploit agent spends its budget discovering the vulnerability instead of validating and exploiting it.

\paragraph{\textbf{Static Analysis Scenario}}
\sysname{} receives all findings produced by Semgrep. The set contains \texttt{FileDownloadController.php:24} as a candidate vulnerability, along with several other plausible findings. The agent must evaluate these candidates, determine which ones correspond to reachable vulnerabilities, and attempt to exploit them. For each finding, it still needs to identify the reachable endpoint, authenticate to the application, construct the request, and verify the resulting effect.

Static analysis helps to resolve the localization barrier, allowing \sysname{} to efficiently test multiple hypotheses. This helps the agent turn potential findings into verified PoEs. In this case, \sysname{} successfully produces a PoE for the target vulnerability using the Semgrep findings. This also demonstrates how \sysname{} can be used to analyze and validate the findings reported by static-analysis tools.

\paragraph{\textbf{Advisory-informed Scenario}}

In this case, \sysname{} receives more semantic information than in the previous condition. It knows that the target contains a path-traversal vulnerability and is given the vulnerable endpoint, but does not know the exact payload required to trigger it. This provides a head start and reduces the burden of exploring the space of possible testing hypotheses. \sysname{} can focus on trying different strategies to determine whether the CVE is reachable and to construct a proof of exploit.

\paragraph{\textbf{Takeaway}}
This case study shows how localization information helps an agent move from a vulnerability hypothesis to a verified PoE. Without prior information, \sysname{} must spend its budget discovering and localizing the vulnerability. The findings of the Static-analysis provide a precise starting point, while the advisory information provides a higher-level description of what to investigate and where to investigate. In both informed settings, \sysname{} uses this information to construct and verify an exploit rather than simply restating the reported vulnerability.

\section{Discussion}

\subsection{Developers and Industry Impact}
For maintainers, the main benefit of a verified PoE is that it reduces the effort required to determine whether a reported vulnerability is real. 
A vulnerability report may identify a vulnerable component or code pattern, but reproducing the issue can require finding the right endpoint, constructing requests, handling authentication, and determining what observable behavior constitutes successful exploitation. 
A PoE packages these steps into a concrete, reproducible test.

The PoE also makes vulnerability reports easier to communicate and validate between teams. 
Rather than asking another developer or security engineer to reproduce a described attack from scratch, the maintainer can provide the executable test together with its success condition. 
Thus, \sysname{} transforms a vulnerability finding into a concrete security test that can be reused for triage, patch validation, and regression testing.

\subsection{Future Work}

Generating verified exploits remains challenging. \sysname{} is currently limited to vulnerabilities that produce an observable server-side effect, and to Day-N vulnerabilities for which a report already exists. 

Future iterations could extend the oracle to further effect classes, such as memory-safety bugs validated through crash or sanitizer oracles. 
Our zero-knowledge condition withholds the advisory from the agent, but still targets a disclosed vulnerability. Whether effect verification can support the discovery of previously undisclosed vulnerabilities remains an open question.

\section{Threats to validity}

Our work is subject to several potential limitations that may influence the interpretation and generalization of the results. In the following, we discuss threats across internal, external, construct, and conclusion validity, highlighting the factors that can compromise robustness and suggesting ways to mitigate them.

\paragraph{\textbf{Construct Validity}}

We consider an exploit successful if we can observe its effect. This observation holds for our eight exploit types. The effect oracle confirms an exploit only after it has observed the effect, so it cannot credit an exploit that did not fire. However, the effect oracle is \emph{mechanism blind}. The oracle only verifies that the effect occurred through a vulnerability, not through a specific vulnerability under test. A different vulnerability yielding the same observable outcome would pass. We mitigate this in the advisory-informed and static-analysis-seeded conditions with a route-and-vector assertion gate that credits an exploit only when the triggering request matches the target CVE's endpoint, injection signature, and authentication precondition. In the zero-knowledge case, no such gate is applied because any genuine exploit is a valid discovery.

\paragraph{\textbf{Internal Validity}}

Post-cutoff disclosure prevents the model from having memorized the specific CVE advisory or proof of concept. It does not prevent the model from having seen the vulnerable source code in pre-cutoff training data, since many of these applications existed before their CVE was filed. The model may also recall well-known older vulnerabilities and find those instead of the target CVE. The agent runs in a sandbox on a separate host, while the application and the verification oracle are hosted separately. In addition, security groups and rules prevent the agent host from communicating directly with the verification oracle. The agent runs long sessions and has stochastic behavior. For OWASP, we report union pass@3. For \benchname{}, we evaluate each condition with one run per target, and manually inspect and re-execute every generated PoE. Since each instance provisions a full application, multiple runs varying planner and exploit models are time consuming and expensive.

\paragraph{\textbf{External Validity}}

Our testing is limited to web applications and eight effects seen on a running server. We run the web applications via Docker containers and provide code access to the agent sandbox. Vulnerabilities with no such effect, like Cross-Site Scripting (XSS) or Denial of Service (DoS), and memory bugs or non-web apps, are not covered. Although an agent may exploit these bugs through its own discovery or through scanning tools, we cannot observe effects for them. 

\paragraph{\textbf{Conclusion Validity}}

For \benchname{}, we report the vulnerability confirmed to reproduce the target CVE. All produced proofs of exploit were manually inspected. The reported OWASP benchmark numbers are union pass@3 to account for transient timeouts or failed API calls.

\section{Related Work}

\paragraph{\textbf{Vulnerability Detection using Static Analysis:}}

Static analysis tools identify potentially vulnerable code by examining source code or bytecode without running the program. Rule-based systems such as \cite{codeql}, \cite{semgrep}, \cite{bandit}, \cite{bearer} rely on taint analysis. Code Property graph methods \cite{yamaguchi2014joern} use unified representations. They use AST, control-flow, and program dependence to track source-to-sink taint flows and vulnerabilities. Although widely used, static analyzers are rarely successful in detecting real-exploitable ones. A comprehensive survey by \citep{hermann2026toolsexploitablevulnerabilitiessurvey} finds that most of the tools concentrate on a narrow range of vulnerabilities. Because taint analysis cannot enumerate every program state, static analyzers produce many false positives and miss runtime behavior that diverges from the analyzed code. 

\paragraph{\textbf{Vulnerability Exploitation using Dynamic Analysis:}}

Dynamic analysis tools (DAST) identify vulnerabilities by running the source code as a live application. Web application scanners such as OWASP ZAP ~\cite{owaspzap} and Burp Suite ~\cite{burpsuite} probe endpoints for injection vulnerabilities. Dynamic analysis creates a Proof of Exploits from successful exploit responses, the same signal on which \sysname{} is based. But  DAST tools operate blindly to application logic and state, driving handcrafted or rule-based payloads at whatever surface they can reach. This limits them to shallow single-request vulnerabilities and ignores any flaw that requires reasoning through several input transformations before a payload reaches its sink. \sysname{} keeps the effect-based confirmation that makes dynamic analysis trustworthy while adding white-box reasoning over the source, letting it construct the multi-step exploits a black-box probe cannot.

\paragraph{\textbf{LLMs for Detecting Vulnerable Code:}}

Recently, Large language models have been applied to enhance vulnerable code detection. IRIS ~\cite{li2025iris} and QLCoder ~\cite{wang2026qlcoder} pair an LLM with static analysis to synthesize better detectors. Combined SAST-and-LLM pipelines report modest recall gains~\cite{zhou2024comparisonstaticapplicationsecurity}. IRIS is bound to a fixed set of CodeQL query templates, QLCoder still carries a high false-positive rate even with a strong backbone model ~\cite{claude-4-6-Sonnet}, and none establishes that a reported flaw can actually be exploited. They advance the detection frontier, but leave the verification gap untouched, and \sysname{} consumes their output as readily as any static analyzer's, converting detected candidates into confirmed exploits.

\paragraph{\textbf{Exploitation Benchmarks:}}

Several benchmarks assess vulnerability detection and exploitation with language models. Each benchmark focuses on a specific challenge. ExploitBench \cite{exploitbench2026} targets the V8 JavaScript and WebAssembly engine bugs commonly found in Chrome, Edge, Node.js, and Cloudflare Workers. CyberGym ~\cite{wang2026cybergym} focuses on memory-safety bugs in C and C++ codebases. It provides 1,507 vulnerability instances sourced from OSS-Fuzz in 188 projects. Agents must generate proof-of-concept tests that trigger crashes to reproduce the bug. CyberGym-E2E ~\cite{shi2026cybergyme2e} extends this to 920 instances that cover the entire lifecycle: vulnerability discovery, PoC generation, and patch generation. It validates four stages, from crash reproduction through functional correctness of the patch. These benchmarks focus on memory-safety issues (buffer overflows, use-after-free, heap corruption). They validate success through crash reproduction or differential testing against patched versions. Our Benchmark \benchname{} covers web application vulnerabilities with observable effects for 8 types, thus extending the landscape from libraries and memory safety issues to full running applications. 

\paragraph{\textbf{Proof of Vulnerabilities Generation:}}

Generating a reliable proof of vulnerability is difficult, as it requires a deep understanding of code flow nested deep through a program. \sysname{} shares its central premise that exploitability is best demonstrated by running something rather than asserting it. FaultLine~\cite{nitin2025faultlineautomatedproofofvulnerabilitygeneration} traces an input from source to sink, explains the branch conditions the input must satisfy, and generates a proof-of-vulnerability test through an iterative feedback loop. PoC Gym ~\cite{gezgin2026pocgymreliablellmassistedproofofconcept} combines static and dynamic information to generate proof-of-concept candidates for CWE-Bench-Java. AnyPoC ~\cite{zhao2026anypocuniversalproofofconcepttest} is a multi-agent framework that synthesizes and independently re-executes proof-of-concept tests for large systems software such as Firefox, Chromium, and SQLite. These methods generate code-level regression or crash-triggering tests, or, in AntiProof's case, focus on detector synthesis over general software. \sysname{} instead verifies an exploit by its observable effect on a live deployed HTTP service, which requires neither a bug oracle nor a patch. 

\paragraph{\textbf{Agent Harness for Security:}}

Autonomous agents have also been applied to offensive security. PentestGPT~\cite{pentestGPT} drives an LLM through interactive black-box penetration testing, and more recent pentesting agents add structure to the search~\cite {deng2026makesgoodllmagent}. Co-RedTeam ~\cite{he2026coredteamorchestratedsecuritydiscovery} coordinates specialized agents for vulnerability discovery and exploitation and, like \sysname{}, grounds its decisions in real execution feedback rather than the model's assertion alone. Model vendors also ship agentic security review within their own products, such as Codex~\cite{codexsec} and Claude Code ~\cite{claudecodesec}, which binds the reviewer to the model and harness of a single provider. We do not evaluate Claude Code Security or Codex Security because their terms of service do not allow the scanning of third-party repositories. The \sysname{} agent loop is open source and runs on any backbone model, consuming the findings of any detector rather than a fixed one, so neither the model nor the vulnerability source is bound to a particular vendor.

\section{Conclusion}

In this paper, we present \sysname{}, a specialized security harness to synthesize verifiable web exploits. Using a hypothesis-graph agent alongside an effect-based oracle, \sysname{} generates Proofs of Exploits (PoEs) that are confirmed by observed effects rather than by the model's own judgment. On \benchname{}, \sysname{} produces verified exploits for $10$ applications and, through coordinated disclosure, reports the first working proof of exploit for $4$ advisories that were reported.

Our evaluation shows that vulnerability localization is a critical factor in real-world exploit coverage. Providing the location of the vulnerability more than doubles the number of applications successfully exploited, from $4$ to $10$ out of $25$. The agent consistently succeeds on exposed, well-localized vulnerabilities, while authenticated and multi-step vulnerabilities remain challenging. These results suggest that progress toward autonomous exploitation will require advances in both vulnerability localization and model capabilities.

\section*{Ethics and Privacy Statement}

\paragraph {\textbf{Evaluation Scope:}}

We evaluated \sysname{} on open-source applications with disclosed vulnerabilities reported through GitHub Security Advisories (GHSA) or the National Vulnerability Database (NVD). These are N-day vulnerabilities, where CVEs are disclosed, but the entire Proof of Exploit is absent. Additionally, we evaluated on the OWASP Benchmark ~\cite{owasp-benchmark-python}, which is synthetic by construction and has a known ground truth.

\paragraph {\textbf{Responsible Disclosure:}}

For advisories that lacked a Proof of Exploit (PoE), \sysname{} generated exploitability proofs. We disclosed every generated exploit to the affected maintainers through coordinated disclosure. \sysname{} operates only on already-disclosed vulnerabilities. We used vulnerabilities disclosed after the models' training cutoff to control for memorization and to evaluate the zero-knowledge condition, in which the agent is given no advisory. \sysname{} was not run to attack production systems or find new vulnerabilities. All targets were local deployments of affected applications.

\paragraph {\textbf{Ethical considerations:}}

We acknowledge that the open source red teaming agent could be repurposed to locate vulnerabilities for malicious purposes. \sysname{} is intended for research and for the review of code that the user owns or is authorized to audit, and we ask that it be used accordingly. The method was created for safety and defense purposes, and we intend no harm to users.

\section*{Data Availability}

Code, execution results, and datasets will be made available soon at \url{https://github.com/LeiLiLab/ulysses-harness}.

\section*{Acknowledgments}
\label{sec:acknowledgments}

This work was conducted as part of the Amazon Nova AI Challenge, which provided funding and AWS compute credits to participating university teams. The study design, model selection, analysis, and the decision to publish were made independently by the authors. We also acknowledge support from national funds through Fundação para a Ciência e a Tecnologia, I.P. (FCT), under projects UID/50021/2025 and UID/PRR/50021/2025. This work is also co-financed by FCT through the Carnegie Mellon Portugal Program under the fellowship PRT/BD/155049/2024. Li is partly supported by the CMU CyLab seed grant.


\clearpage
\printbibliography

\newpage
\appendix
\newpage

\section{Appendix}
\label{sec:appendix}

\begin{table*}[h]
    \caption{Vulnerability instances included in our benchmark.}
    \label{tab:instances}
    \begin{tabular}{lllllrr}
      \toprule
      \textbf{Repository} & \textbf{CVE} & \textbf{Disclosed} & \textbf{Lang.} & \textbf{CWE} & \textbf{Weakness Class} & \textbf{kLoC} \\
      \midrule
      atlantis      & CVE-2026-64679 & Aug'26 & Go   &  22 & Path Traversal         &   72.5 \\
      filebrowser   & CVE-2026-35585 & Apr'26 & Go   &  78 & OS Command Injection   &   21.7 \\
      seaweedfs     & CVE-2026-72920 & Aug'26 & Go   & 306 & Missing Authentication &  644.6 \\
      sftpgo        & CVE-2026-49244 & Aug'26 & Go   &  22 & Path Traversal         &  176.9 \\
      siyuan        & CVE-2026-72811 & Sep'26 & Go   &  89 & SQL Injection          &  291.0 \\
      upsnap        & CVE-2026-49819 & Aug'26 & Go   & 306 & Missing Authentication &    4.8 \\
      \addlinespace
      databasir     & CVE-2025-66944 & Mar'26 & Java &  89 & SQL Injection          &   41.2 \\
      rillflow      & CVE-2025-4866  & May'25 & Java &  74 & Injection              &   85.6 \\
      yamcs         & CVE-2026-44632 & Jul'26 & Java &  94 & Code Injection         & 1150.5 \\
      \addlinespace
      fuxa          & CVE-2026-47720 & Aug'26 & JS   &  89 & SQL Injection          &   84.2 \\
      \addlinespace
      basercms      & CVE-2026-30940 & Mar'26 & PHP  &  22 & Path Traversal         &  279.6 \\
      churchcrm     & CVE-2025-1023  & Feb'25 & PHP  &  89 & SQL Injection          &  212.7 \\
      dolibarr      & CVE-2026-22666 & Apr'26 & PHP  &  95 & Eval Injection         & 1949.7 \\
      freshrss      & CVE-2025-58173 & Dec'25 & PHP  &  20 & Input Validation       &  110.5 \\
      mybb          & CVE-2026-45117 & Aug'26 & PHP  &  94 & Code Injection         &  199.3 \\
      openemr       & CVE-2026-40506 & Aug'26 & PHP  &  22 & Path Traversal         & 1325.4 \\
      photoshow     & CVE-2023-53981 & Dec'25 & PHP  &  78 & OS Command Injection   &   25.6 \\
      rconfig       & CVE-2026-77914 & Aug'26 & PHP  &  22 & Path Traversal         &  230.8 \\
      wintercms     & CVE-2026-63179 & Aug'26 & PHP  &  22 & Path Traversal         &  208.0 \\
      \addlinespace
      astrbot       & CVE-2025-48957 & Jun'25 & Python &  23 & Path Traversal         &   44.0 \\
      crawl4ai      & CVE-2026-56262 & Jun'26 & Python & 306 & Missing Authentication &  214.8 \\
      fireshare     & CVE-2026-33645 & Mar'26 & Python &  22 & Path Traversal         &   16.5 \\
      frigate       & CVE-2026-25643 & Feb'26 & Python &  78 & OS Command Injection   &  104.5 \\
      langflow      & CVE-2025-3248  & Apr'25 & Python & 306 & Missing Authentication &  184.9 \\
      openhands     & CVE-2026-33718 & Mar'26 & Python &  78 & OS Command Injection   &  327.1 \\
      \bottomrule
    \end{tabular}
\end{table*}


\end{document}